\documentclass[aps,pra,reprint,showpacs,nofootinbib,superscriptaddress]{revtex4-1}
\usepackage{dcolumn}   
\usepackage{bm}        
\usepackage{amssymb}   
\usepackage{mathtools}
\usepackage{amsthm}
\usepackage{ragged2e}
\usepackage{verbatim}
\usepackage[colorlinks=true,linkcolor=blue,citecolor=red,plainpages=false,pdfpagelabels]{hyperref}
\usepackage{color,xcolor,colortbl}
\usepackage{multirow}
\usepackage{boxedminipage}
\usepackage{bbm}
\usepackage{float}
\usepackage{array}

{\rm }

\def\be{\begin{equation}}
 \def\ee{\end{equation}}
 \def\bea{\begin{eqnarray}}
 \def\eea{\end{eqnarray}}
 \def\bes{\begin{eqnarray}}
 \def\ees{\end{eqnarray}}
 \def\bi{\begin{itemize}}
 \def\ei{\end{itemize}} 

\def\2{\frac{1}{2}}
\def\4{\frac{1}{4}}

\def\dis{\operatorname{dis}}

\newcommand{\bra}[1]{\langle #1|}
\newcommand{\ket}[1]{|#1\rangle}
\newcommand{\tn}[1]{\textsf{#1}}

\begin{document}

\title{Practically secure quantum position verification}

\author{Siddhartha Das}\email{sidddas@ulb.ac.be} \affiliation{Hearne Institute for Theoretical Physics, Department of Physics and Astronomy, Louisiana State University, Baton Rouge, Louisiana, 70803, USA}
\affiliation{Centre for Quantum Information \& Communication (QuIC), \'{E}cole polytechnique de Bruxelles,   Universit\'{e} libre de Bruxelles, Brussels, B-1050, Belgium}

%
%
\author{George Siopsis} \email{siopsis@tennessee.edu}\affiliation{Department of Physics and Astronomy, The University of Tennessee, Knoxville, Tennessee, 37996-1200, USA}

\date{\today}

\begin{abstract}

	We discuss quantum position verification (QPV) protocols in which the verifiers create and send single-qubit states to the prover. QPV protocols using single-qubit states are known to be insecure against adversaries that share a small number of entangled qubits. We introduce QPV protocols that are \emph{practically} secure: they only require single-qubit states from each of the verifiers, yet their security is broken if the adversaries  sharing an \emph{impractically} large number of entangled qubits employ teleportation-based attacks. These protocols are a modification of known QPV protocols in which we include a classical random oracle without altering the amount of quantum resources needed by the verifiers. We present a cheating strategy that requires a number of entangled qubits shared among the adversaries that grows exponentially with the size of the classical input of the random oracle.

\end{abstract}

\maketitle

\section{Introduction}

	Suppose that a security organization would like to identify the position of its spy in a secure location, who could possibly be surrounded by adversaries, before initiating any distant private communication. The security organization could execute a protocol whose task is to use the spatial position of the spy as its only credential that has to be verified by the organization. In general, there are situations, such as in position-based cryptography \cite{BC93,CGMO09}, in which it is in the interest of the collaborating parties to authenticate their positions before initiating any secure communication. Protocols to achieve such a task are often called position verification. 

	In the task of position verification, we assume that a prover $P$ is located at a fixed spatial position \textsf{pos}. There is a set $\{V_i\}_{i=0}^{K-1}$ of $K$ verifiers located at different positions. A time-bound interactive protocol is allowed to run between the verifiers and the prover in order for the prover to convince the verifiers of his position credential. All the verifiers can communicate privately among themselves and collectively agree on items that each individual verifier would send to the prover along with the task that the prover has to perform. The prover is expected to send the information obtained at the end of the task performed on the received items within the time limit set by the verifiers, which is typically equal to the time that a signal would take to travel from \textsf{pos} to the farthest verifier.

	In this paper, we discuss quantum position verification (QPV) protocols in which the verifiers and the prover employ quantum strategies against adversaries capable of quantum attacks. Our goal is to develop QPV schemes that are practically secure: while the verifiers use a few qubits, the adversaries need an exponentially large amount of resources (shared entanglement and quantum computational power) to break the security of the protocol. The important point to note is that the entanglement distribution over long distances and the storage of entangled qubits are technologically challenging (\emph{cf}.~with Refs.~\cite{DKD17,SAA+10}) which limit the attacking capability of the adversaries. In this sense, we state that the schemes we have presented are technologically feasible and practically secure.

	In 1993, Brands and Chaum \cite{BC93} introduced the ``distance bounding'' technique  in the classical setting by timing the delay between sending out of a challenge bit from a verifier to the prover and receiving back the corresponding response bit. If the speed of communication is bounded by the speed of light, this technique gives an upper bound on the distance between the prover and the verifier. These ideas were extended in Ref.~\cite{CGMO09} (in the classical setting) to what is now known as position verification. In particular, in the so-called ``Vanilla model'', the prover is located at a position \textsf{pos} that lies inside the tetrahedron enclosed by the verifiers. In this model, there is always a possibility that a group of adversaries can collectively disguise themselves as the honest prover by convincing the verifiers of being located at \textsf{pos} even when they are all positioned elsewhere. It is assumed that an adversary can locally store all information she receives, and at the same time share this information with other colluding adversaries located elsewhere. This impossibility result rules out the existence of a secure position verification protocol under classical settings even when one makes computational hardness assumptions on the adversaries \cite{CGMO09,BCF+14}. 

	A natural question that arises is whether there exists a secure position verification protocol in the quantum setting. One early position-based cryptography protocol in the quantum setting is quantum tagging, first discussed in 2002 and described in a 2006 patent, Ref.~\cite{KMSB06}. Quantum tagging is the task of authenticating the location of a classical tagging device by sending and receiving quantum signals from distant sites. It is assumed that adversaries control the environment, and that their quantum information processing and transmitting power is unbounded. In Ref.~\cite{KMS11}, several schemes for the quantum tagging task were described, and their security breach using quantum-teleportation-based attacks were discussed. 

   After the introduction of quantum tagging~\cite{KMSB06}, a few other proposals of secure QPV protocols were discussed in Refs.~\cite{Mal10a,Mal10b,BCF+14} \footnote{The results of the US patent \cite{KMSB06} appeared in publicly accessible scientific literature in August 2010 \cite{KMS11}, whereas, \cite{Mal10a} appeared in March 2010, \cite{Mal10b} appeared in April 2010, and \cite{BCF+14} appeared in August 2010.}. The possibility of ``instantaneous measurement" of non-local variables (observables)~\cite{Vai03} leads to the breaking down of security of such QPV protocols by colluding adversaries performing teleportation-based attacks \cite{BCF+14,BK11}. As it turns out, all of these proposed schemes can be broken by colluding adversaries employing teleportation-based attacks \cite{KMS11,BCF+14,LL11,BFSS13}. Various other QPV protocols and attack strategies by adversaries have been proposed \cite{TFKW13,BFSS13,Unr14,CL15,QS15,Spe16,LXS+16,RMW16} along with a security analysis with different physical constraints on the colluding adversaries, among which Ref.~\cite{Unr14} was the first to make use of random oracle in the protocol. Some works have also established lower bounds on the number of entangled pairs required by the adversaries to breach the security of certain QPV protocols \cite{BK11,TFKW13,BFSS13,RG15,Unr14}.  
   
   We note, however, that others have argued that quantum location verification protocols do exist for which no undetectable attack is known \cite{Malaney2016}.

	An important feature of a QPV protocol is the limited time in which the prover can perform the computations and communicate with the verifiers. This suggests the possibility of strengthening QPV protocols by taking this time limit into account and negating the practical feasibility of teleportation-based attacks by colluding adversaries within the given time limit. As a consequence of such limited-time constraints, it is possible that the adversaries would need to share a very large amount of resources (entangled pairs and quantum computational power) between them for each round of the protocol to breach security, while the verifiers would use only a few low-dimensional quantum states for the protocol.

	In this work, we introduce one such protocol by modifying previously-known protocols. Our protocol uses single-qubit states from each verifier and makes use of a classical random oracle held by the verifiers and the prover. By using the best-known teleportation-based attack strategies \cite{BCF+14,BK11}, we show that the number of entangled qubits that need to be shared among the adversaries in order to breach security grows exponentially with the size of the classical input of the random oracle. In Section~\ref{sec:QPV}, we introduce QPV and fix the notation. In Section~\ref{sec:rev-old-p}, we present known QPV protocols that make use of single-qubit states. As a warm-up to the next Section, we also introduce a modification by adding classical information into the protocols which tightens their security. In Section~\ref{sec:new-P}, we introduce novel QPV protocols by adding a classical random oracle. These protocols appear to be practically secure under the attack of colluding adversaries sharing a large amount of entangled pairs (exponentially growing with the length of classical information), even though each verifier sends just one qubit to the prover to execute the QPV protocol.  Finally, in Section~\ref{sec:con}, we conclude.

\section{Quantum position verification}\label{sec:QPV}

	The goal of a QPV protocol is for a set $\tn{Ver}(K)=\{V_i\}_i$, $i\in\{0,1,\dotsc,K-1\}$, of $K$ verifiers to authenticate the spatial position $\textsf{pos}$ of a prover $P$. The spatial positions of all the parties involved are fixed in time. The prover $P$ is assumed to lie within the convex hull formed from the spatial positions of the verifiers. For all $i$, let $\textsf{pos}_i$ denote the spatial position of $V_i$. All the verifiers can securely communicate among themselves to decide on a list $\tn{Item}=\{\tn{item}_i\}_i$ of items, where $\tn{item}_i$ corresponds to items that are transmitted from $V_i$ to $\textsf{pos}$. Each $\tn{item}_i$ comprises arrays of classical bits and quantum states. The verifiers and the prover agree upon the set $\tn{Opn}$ of operations that the prover has to perform based on the elements of $\tn{Item}$. All the measurement operations and computations by $P$ are assumed to be instantaneous. The result $\tn{Rslt}$ at the end of operations instructed in $\tn{Opn}$ is broadcast to all the verifiers. The information communication, $\tn{Item}$ and $\tn{Rslt}$, between the prover and the verifiers is assumed to take place at the speed $c$ of light. For simplicity, we set $c=1$. Then the time taken for information to travel between the verifiers and the prover is equal to the spatial distance between them.

	Now, suppose that there is a set $\tn{Adv}=\{E_i\}_i$ of colluding adversaries, each $E_i$ positioned at $\textsf{pos}^\prime_i$ between $V_i$ and $\textsf{pos}$. These adversaries want to cheat the verifiers by convincing them of being positioned at $\textsf{pos}$, even though $\textsf{pos}^\prime_i\neq \textsf{pos}$ for all $i$. We restrict the adversaries to make use of resources available only at their positions $\tn{pos}^\prime_i$ for all $i$. However, they may share non-local quantum resources, such as entanglement. They are allowed to collude through classical communication among each other. Classical communication among the adversaries, and between them and the verifiers, is assumed to be at the speed of light. We denote the Euclidean distance between any two spatial positions $\textsf{pos}_i$ and $\textsf{pos}$ by $\dis(\textsf{pos}_i,\textsf{pos})$. Furthermore, we assume that the spatial distance between $V_i$ and $\textsf{pos}$ is the same, i.e., $\dis(\textsf{pos}_i,\textsf{pos})=d$. If the verifiers transmit the information $\{\tn{Item},\tn{Opn}\}$ at time $t_\ell$ towards the prover, then the result $\tn{Rslt}$ has to arrive back to the verifers at time $t_\ell+2d$. The verfiers accept the prover's position credential only when the expected result is received from the prover on time. Let us denote this $1-$round scheme as $\tn{QPV}[\tn{Ver}(K),\tn{P}(\textsf{pos}),\tn{Item},\tn{Opn},\tn{Rslt},d]$. 

	The security of a generic QPV protocol is generally analyzed using the completeness and soundness conditions \cite{CGMO09,BCF+14}. $\tn{QPV}[\tn{Ver(K)},\tn{P}(\textsf{pos}),\tn{Item},\tn{Opn},\tn{Rslt},d]$ is said to have perfect completeness if the verifiers always agree with an honest prover $P$. In other words, if the verifiers accept the prover's position credential of being spatially located at $\textsl{pos}$ with probability $1$, then the protocol is said to have perfect completeness. $\tn{QPV}[\tn{Ver(K)},\tn{P}(\textsf{pos}),\tn{Item},\tn{Opn},\tn{Rslt},d]$ is said to be $\varepsilon$-sound if for any coalition of adversaries $\{E_i\}_i$ spatially located at $\textsf{pos}^\prime_i\neq \textsf{pos}$ for all $i$ and limited to resources $\tn{Res}$ that are only locally available at these positions, the verifiers accept with probability at most $\varepsilon$.

\begin{figure}[H]
		\centering
		\includegraphics[scale=1]{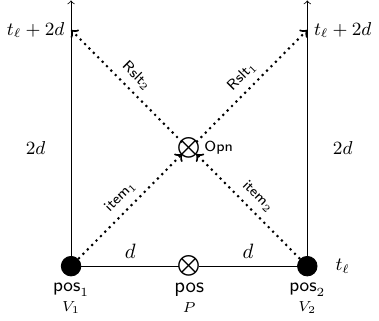}
		\caption{Schematic diagram for the 1-D QPV protocol $\tn{QPV}[\tn{Ver(2),P(\textsf{pos}),\tn{Item},\tn{Opn},\tn{Rslt}},d]$.}\label{fig-qpv}
	\end{figure}

	For our discussion, we consider quantum position verification protocols $\tn{QPV}[\tn{Ver(K),P(\textsf{pos}),\tn{Item},\tn{Opn},\tn{Rslt}},d]$ in one dimension (1-D). We constrain our discussion to qubit systems. For simplicity of discussion, we assume the protocol to have perfect completeness. For the 1-D case, it is sufficient to let $K=2$, so that we have two verifiers, $V_0$ and $V_1$, spatially positioned at the two ends of a line, and a prover $P$ at the middle of the line denoted $\textsf{pos}$, see Fig.~\ref{fig-qpv}. The spatial distance between $V_0$ and $V_1$ is $2d$. The verifiers wish to verify that $P$ is spatially located at $\textsf{pos}$. Unfortunately, there are two adversaries, $E_0$ (between $V_0$ and $P$) and $E_1$ (between $V_1$ and $P$) who will try to fake $P$. Given the geometrical setting of the verifiers and the prover, the optimal number of adversaries required to analyze the security of the protocol is equal to the number of verifiers.

	Let us denote the computational basis of a qubit system by $\{\ket{0},\ket{1}\}$. A two-qubit system in the state
		\be\label{eqBell} |\Phi^+ \rangle = \frac{1}{\sqrt{2}} \left( |00\rangle + |11\rangle \right) \ee
will be referred to as an EPR pair. Any two-qubit maximally entangled state is unitarily equivalent to an EPR pair. The four Bell states are
	\begin{align}\label{eq:Bell}
		\ket{\Phi^{\pm}}&=\frac{1}{\sqrt{2}}(\ket{00}\pm\ket{11}),\nonumber\\
		\ket{\Psi^{\pm}}&=\frac{1}{\sqrt{2}}(\ket{01}\pm\ket{10}).
	\end{align}
	The states $\{\ket{\Phi^+},\ket{\Phi^-},\ket{\Psi^+},\ket{\Psi^-}\}$ form an orthonormal basis for the Hilbert space of two-qubit systems. A Bell measurement is defined to be a projective measurement in this basis. We denote the qubit Hadamard transform by $H$, which is defined by
	\begin{align}
		H\ket{0}&=\frac{1}{\sqrt{2}}(\ket{0}+\ket{1}),\\
		H\ket{1}&=\frac{1}{\sqrt{2}}(\ket{0}-\ket{1}).
	\end{align}
	We let $X$, $Y$, and $Z$ denote the Pauli operators, where $Z=\ket{0}\bra{0}-\ket{1}\bra{1}$ and $X=HZH$ \cite{NC00}.

\subsection{Protocols with single-qubit states}\label{sec:rev-old-p}

	In this section, we discuss some known 1-D QPV protocols $\tn{QPV}[\tn{Ver(2),P(\textsf{pos}),\tn{Item},\tn{Opn},\tn{Rslt}},d]$ and their security breaches that have been discussed in prior work \cite{KMS11,CGMO09,LL11,Mal10a,TFKW13,Unr14}. For the breach of security, a single EPR pair shared among the adversaries suffices. Then, as a warm-up to the implementation of the classical random oracle in the Section~\ref{sec:P-I''}, we modify these protocols by introducing a single bit of information, and discuss the effect on the quantum resources of the adversaries that are needed for breach of security.

\subsubsection{One-qubit protocol}\label{sec:P-I}

	This protocol has been inspired by the Bennett-Brassard 1984 (BB84) quantum-key-distribution (QKD) protocol \cite{BB84}.

	To verify the position of $P$,  the following scheme is employed:
	\begin{enumerate}
		\item The verifiers agree on random bits $x, \theta \in \{ 0,1 \}$. $V_0$ prepares a qubit in the state
			\be |\psi \rangle = H^\theta |x\rangle \ee
		and sends it to $P$. $V_1$ sends $\theta$ to $P$, so they arrive at $P$ at the same time. That is, $\tn{item}_0=\{|\psi\rangle\}$ and $\tn{item}_1=\{\theta\}$. 
		\item As soon as $|\psi\rangle$ and $\theta$ arrive, $P$ performs a measurement in the basis $\{ H^\theta |0\rangle , H^\theta |1\rangle \}$,  and sends the outcome $x^\prime$ to both $V_0$ and $V_1$. The given measurement consititutes $\tn{Opn}$ and $x^\prime$ is the $\tn{Rst}$.  \label{p-1-s-2}
		\item If the verifiers receive $x'$ at the time consistent with the position of $P$, \emph{and} $x'=x$, then they accept; otherwise they reject.
	\end{enumerate}
	Now, suppose there are two adversaries, $E_0$ (between $V_0$ and $P$) and $E_1$ (between $V_1$ and $P$). Can they fake $P$? 
Suppose that the adversaries share no entangled qubits, although they may have qubits in their possession. When $V_0$ sends $|\psi\rangle$, $E_0$ will intercept it before $P$ does, but will not be able to do Step \ref{p-1-s-2}, because she will have to wait for $\theta$ to arrive first. By the time it arrives, it will be too late to send any information to $V_1$ (but not to $V_0$).

	The best strategy is the following, which is based on minimum-error state discrimination \cite{Hel69,Hel76book}. When $E_0$ receives $|\psi\rangle$ from $V_0$, she performs a measurement in the basis
	\be |0'\rangle = \cos \frac{\pi}{8} |0\rangle + \sin\frac{\pi}{8} |1\rangle \ \ , \ \ \ \
|1'\rangle = - \sin\frac{\pi}{8} |0\rangle + \cos\frac{\pi}{8} |1\rangle \ee
and sends the outcome of the measurement to $E_1$. The probability of success of this optimal strategy is
	\be\label{eq5} \epsilon = \cos^2 \frac{\pi}{8} = \frac{1}{2} + \frac{1}{2\sqrt{2}}  \approx 0.85 \ee
	As the above is repeated $n$ times, the success probability becomes $\epsilon^n$ (exponentially small). Notice that the adversaries are not able to make use of the classical information $\theta$ available to them. This information would have been useful, had $E_0$ received it before deciding on what measurement to perform.

	However, $E_0$ can make use of $\theta$ when deciding on the measurement, if the adversaries share entangled qubits, because of the possibility of teleportation. Suppose the adversaries share an EPR pair. Then they can fake $P$ following these steps.
	\begin{enumerate}
	\item Upon receiving $|\psi\rangle$, $E_0$ teleports it to $E_1$. In doing so, $E_0$ performs Bell measurements with outcome $k = \overline{k_0k_1}$, in binary notation. They determine the state $E_1$ receives (instantaneously) as
\be\label{eq:Bellmod} |\phi_k\rangle = X^{k_0} Z^{k_1} |\psi\rangle \ee
Since $|\psi\rangle = H^\theta |x\rangle$, for $\theta = 0$, $|\phi_k\rangle$ is an eigenstate of $Z$, whereas for $\theta =1$, $|\phi_k\rangle$ is an eigenstate of $X = HZH$. We easily obtain
\be H^\theta Z H^\theta |\phi_k\rangle = (-1)^{x\oplus k_\theta} |\phi_k\rangle \ee
$E_0$ sends the results $k$ of her Bell measurements to $E_1$.
	\item At the same time, knowing $\theta$ (having received it from $V_1$), $E_1$ measures $H^\theta Z H^\theta$ (i.e., $Z$, if $\theta =0$, and $X$, if $\theta =1$), and obtains outcome $(-1)^{x\oplus k_\theta}$. She immediately sends both $\theta$ and $(-1)^{x\oplus k_\theta}$ to $E_0$.
	\item Upon receiving the classical information $\{ \theta , (-1)^{x\oplus k_\theta} \}$ from $E_1$, $E_0$ multiplies $(-1)^{x\oplus k_\theta} (-1)^{k_\theta} = (-1)^x$ to calculate $x$ and send the information to $V_0$.
\item  Upon receiving the classical information $k$ from $E_0$, $E_1$ multiplies $(-1)^{x\oplus k_\theta} (-1)^{k_\theta} = (-1)^x$ to calculate $x$ and send the information to $V_1$.
	\end{enumerate}
	This strategy has 100\% probability of success for the adversaries. Thus a single EPR pair among the adversaries is sufficient for breach of security.


\subsubsection{Modified one-qubit protocol}\label{sec:P-I'}

	The above conclusion on breach of security can be avoided by upgrading $|\psi\rangle$ to a $n$-qubit state  with $n>1$. In this case, if the adversaries share $m$ EPR pairs, then the probability of success for the adversaries is given by $\epsilon \le 2^m \cos^{2n} \frac{\pi}{8}$, where the additional factor $2^m$ is due to the availability of the Hilbert space of the entangled pairs. However, realizing such a protocol with multi-qubit states $|\psi\rangle$ is experimentally challenging. Instead, we can modify the above protocol by introducing an additional classical bit of information. In the modified protocol, to verify the position of $P$,  the following scheme is used.
	\begin{enumerate}
		\item The verifiers agree on random $x, \theta_0, \theta_1 \in \{ 0,1 \}$. $V_0$ prepares a qubit in the state
			\be |\psi \rangle = H^{\theta_0\cdot \theta_1} |x\rangle \ee
			and sends it to $P$, along with $\theta_0$. $V_1$ sends $\theta_1$ to $P$, so they arrive at $P$ at the same time.
		\item As soon as $|\psi\rangle$ and $\theta_0,\theta_1$ arrive, $P$ computes (classically) $\theta = \theta_0\cdot\theta_1$, performs a measurement in the basis $\{ H^\theta |0\rangle , H^\theta |1\rangle \}$, and sends the outcome $x'$ to both $V_0$ and $V_1$.
		\item If the verifiers receive $x'$ at the time consistent with the position of $P$, \emph{and} $x'=x$, then they accept; otherwise they reject.
	\end{enumerate}

	Unlike the previous protocol, adversaries with a prior single pair of entangled qubits will not be able to break the security of this modified protocol. This is because $E_1$ has insufficient information to perform the correct measurement on her qubit. $E_1$ can optimize her measurement, but the adversaries can never achieve a 100\% success rate. The adversaries need at least 2 entangled pairs.

Suppose that the adversaries share two EPR pairs, labeled $0$ and $1$, each in the Bell state \eqref{eqBell}.
Then they can fake $P$ following these steps.
\begin{enumerate}
	\item Upon receiving $|\psi\rangle$ and $\theta_0$, $E_0$ teleports $|\psi\rangle$ to $E_1$ using the EPR pair labeled $\theta_0$. In doing so, $E_0$ performs a Bell measurement with outcome $k = \overline{k_0k_1}$, in binary notation. The state $E_1$ receives (instantaneously) is given by \eqref{eq:Bellmod}.
	$E_0$ sends the results $k,\theta_0$ of her Bell measurement to $E_1$. 
	\item At the same time, knowing $\theta_1$ (having received it from $V_1$), $E_1$ measures $Z$ on qubit $0$ and $H^{\theta_1} Z H^{\theta_1}$ (i.e., $Z$, if $\theta_1 =0$, and $X$, if $\theta_1 =1$) on qubit $1$. She obtains outcome $(-1)^{x\oplus k_\theta}$ on the qubit belonging to the EPR pair $E_0$ used to teleport $\ket{\psi}$, and $\lambda$ on the other qubit. She immediately sends both $\theta_1$ and $((-1)^{x\oplus k_\theta}, \lambda)$ to $E_0$.
	\item Upon receiving the classical information $\{ \theta_1 , ((-1)^{x\oplus k_\theta}, \lambda) \}$ from $E_1$, $E_0$ multiplies $(-1)^{x\oplus k_\theta} (-1)^{k_\theta} = (-1)^x$ to calculate $x$ and send the information to $V_0$. She knows which of the two outcomes $\lambda$ is, because that is determined by $\theta_0$.
	\item  Upon receiving the classical information $k$ and $\theta_0$ from $E_0$, $E_1$ multiplies $(-1)^{x\oplus k_\theta} (-1)^{k_\theta} = (-1)^x$ to calculate $x$ and send the information to $V_1$. Again, she knows which of the two outcomes $\lambda$ is, because that is determined by $\theta_0$.
\end{enumerate}
This strategy has 100\% probability of success for the adversaries.

\subsubsection{Two-qubit protocol}\label{sec:P-II}

	This is a scheme making use of entanglement of two qubits received by the prover, and is secure if the adversaries share no EPR pairs \cite{LXS+16}.
	
	To verify the position of $P$,  the following scheme making use of two qubits is employed:
	\begin{enumerate}
		\item The verifiers agree on random $x_0, x_1, \theta \in \{ 0,1 \}$. $V_i$ prepares a qubit in the state $H^\theta |x_i\rangle$ ($i=0,1$) and sends it to $P$. Both states arrive at $P$ at the same time.
		\item $P$ performs a measurement projecting onto the state $\ket{\Psi^+}$ \eqref{eq:Bell}. If the measurement is successful, then he sends $z=1$, otherwise he sends $z=0$.
		\item The verifiers accept if the result $z$ of $P$'s measurement is consistent with the states sent by them to $P$. The verifiers receive $z=1$ half of the time, if they send the same state with $\theta=0$ (different states with $\theta=1$), and always $z=0$ if they send different states with $\theta=0$ (same state with $\theta=1$).
	\end{enumerate}
It should be noted that it is advantageous for the verifiers if $P$ projects onto both $\ket{\Psi^+}$ and $\ket{\Psi^-}$, making it harder for adversaries to mimic $P$'s actions. It is straightforward to extend the analysis presented here to this case. For simplicity, we omit the discussion. 

For the security analysis, first let us consider the case when the adversaries do not share any entangled pairs. $E_0$ intercepts the qubit from $V_0$ and measures it in the $\{ |\hat{\mathbf{n}}_1\rangle, |\hat{\mathbf{n}}_2\rangle \}$ basis, where $\hat{\mathbf{n}}_1\perp \hat{\mathbf{n}}_2$. Similarly, $E_1$ intercepts the qubit from $V_1$ and measures it in the same basis. They communicate their results to each other. If they disagree, they report $z=1$ to the verifiers half of the time. If they agree, they report $z=0$ to the verifiers.

The probability of error for the adversaries is
\bea P_{\text{err}} &=& \frac{1}{4} \sum_{\theta=0}^{1}\left( |\langle 0 |H^\theta|\hat{\mathbf{n}}_1\rangle |^2 + |\langle 1|H^\theta |\hat{\mathbf{n}}_2\rangle |^2 \right) \nonumber\\
&&\ \ \ \ \ \ \ \ \times\left( |\langle 0|H^\theta |\hat{\mathbf{n}}_2\rangle |^2 + | \langle 1|H^\theta |\hat{\mathbf{n}}_1\rangle |^2 \right) 
\eea
Let $|\hat{\mathbf{n}}_1\rangle = \left( \alpha , \beta \right)^T$, $|\alpha|^2+|\beta|^2=1$, and $|\hat{\mathbf{n}}_2\rangle = \left( -\beta^\ast , \alpha^\ast \right)^T$. Then
\be P_{\text{err}} = |\alpha \beta|^2  + \frac{1}{4} |\alpha + \beta |^2 |\alpha - \beta |^2 
\ee
It is easy to see that
\be P_{\text{err}} = \frac{1}{4} + |\Im\alpha^\ast \beta|^2 \ge \frac{1}{4}.\ee
The error is minimized when, e.g., $\alpha = 1$, $\beta = 0$. Correspondingly, the probability of success is
\be\label{eq15} \epsilon = 1 - P_{\text{err}} \le \frac{3}{4}~. \ee
This bound compares favorably to the result \eqref{eq5} for the single-qubit protocol.

Next, suppose that the adversaries share a pair of qubits in the state \eqref{eqBell}. Once $E_0$ intercepts the qubit from $V_0$, she can perform a Bell measurement on her qubit in the pair shared with $E_1$ and the intercepted qubit, projecting it onto one of the orthogonal states $\{ |\Phi^+\rangle, |\Phi^-\rangle, |\Psi^+\rangle, |\Psi^-\rangle \}$ (assuming unlimited technological capabilities). $E_1$ performs a similar measurement on her half of the pair and the qubit she intercepts from $V_1$. They report the results to each other. They send $z=1$ to the verifiers, if $E_0$ measures $|\Psi^+\rangle$ and $E_1$ measures $\ket{\Phi^+}$.

Since
\be _{V_0E_0}\langle \Psi^+| {}_{V_1E_1}\langle \Phi^+ |\Phi^+ \rangle_{E_0E_1} = {}_{V_0V_1}\langle \Psi^+ | \ee
their operation is equivalent to the prover's measurement, and therefore they have 100\% probability of success.

\subsubsection{Modified two-qubit protocol}\label{sec:P-IIa}

Let us introduce two classical bits of information into the protocol, similar to the modified one-qubit protocol. The steps of the protocol are as follows:
\begin{enumerate}
	\item The verifiers agree on random $x_0, x_1, \theta, y_0, y_1 \in \{ 0,1 \}$. $V_i$ prepares a qubit in the state
	$H^\theta |x_i\rangle$ and sends it to $P$, along with ${y}_i$ ($i\in\{0,1\}$). Here, $\tn{itm}_i=\{{y}_i,H^\theta |x_i\rangle\}$, for $i\in\{0,1\}$. Both states arrive at $P$ at the same time.
	\item\label{step:2} $P$ computes (classically) $y = y_0\cdot y_1$, and applies $H^y$ to each of the states he receives. Then he performs a measurement projecting onto the state $\ket{\Psi^+}$ \eqref{eq:Bell}, and signals $z=1$ or $0$ to the verifiers, depending on whether his measurement was successful or not.
\end{enumerate}
Without prior shared entanglement, the adversaries have a success probability given by \eqref{eq15}, as before. However, they are no longer able to take advantage of a single shared EPR pair, because they do not have the classical information needed to mimic step \ref{step:2}, and perform the correct Bell measurements. Therefore, they need at least two shared EPR pairs for a security breach.

It appears that the adversaries need a larger number of entangled pairs. Suppose that the adversaries share $5$ maximally entangled pairs labeled as ${a}\in\{0,1,2,3,4 \}$, each in the Bell state \eqref{eqBell}, so that the state of the system shared between the adversaries is the tensor product $|\Phi^+\rangle_1 |\Phi^+\rangle_2 |\Phi^+\rangle_3 |\Phi^+\rangle_4 |\Phi^+\rangle_5$. The adversaries will use them in a complex scheme involving teleportation to fake $P$. Here are the steps involving pairs with labels as indicated:
\begin{enumerate}
	\item Upon receiving $H^\theta \ket{x_0}$ and ${y}_0$, $E_0$  teleports the state to $E_1$ using the EPR pair labeled ${a}=0$. In doing so, $E_0$ performs a Bell measurement, and $E_1$ receives the state
	\be\label{eq:Bellmod2}  X^{k_0} Z^{k_1} H^\theta \ket{x_0} ~. \ee She also sends the classical information $k=\overline{k_0k_1}$, as well as ${y}_0$ to $E_1$.
	\item Upon receiving $H^\theta \ket{x_1}$ and ${y}_1$, $E_1$  teleports the state to $E_0$ using the EPR pair labeled ${a}=2y_1+1$. $E_0$ receives
	\be\label{eq:Bellmod2b} X^{k_0'} Z^{k_1'} H^\theta \ket{x_1} ~. \ee
	She also teleports back to $E_0$ the state \eqref{eq:Bellmod2} using the EPR pair labeled $a=2y_1+2$. 
		Thus, $E_0$ receives the state $X^{k_0+k_0''} Z^{k_1+k_1''} H^\theta \ket{x_0}$, which can be simplified, if $E_0$ applies $X^{k_0} Z^{k_1}$ (since $k$ is known to $E_0$) to 
	\be X^{k_0''} Z^{k_1''} H^\theta \ket{x_0}~. \ee She also sends the classical information ${y}_1$ to $E_0$.
	\item $E_0$ applies $H^{y_0}$ to the channels $a=3,4$, thus effectively applying $H^{y}$ ($y=y_0\cdot y_1$) to the states she received from $E_1$. She then performs a Bell measurement on each of the pairs labeled $(1,2)$ and $(3,4)$. In each case, she reports success to $E_1$, if the outcome is $\ket{\Psi^+}$ \eqref{eq:Bell}.
	\item Upon receiving $y_0$ and the ``success" report from $E_0$, $E_1$ reports $z=0$ or $1$ to $V_1$, accordingly, knowing which pair of channels contains the teleported states. At the same time, upon receiving $y_1$ from $E_1$, $E_0$ learns the pair of channels containing the teleported states, and reports $z=0$ or $1$, accordingly, to $V_0$. 
\end{enumerate}
The above protocol can only succeed if $k' = k'' =0$, which occurs with probability $\frac{1}{16}$. The adversaries can increase their odds at the expense of adding EPR pairs. A large number of EPR pairs are needed for 100 \% success rate \cite{BCF+14}.

\section{Protocols with single-qubit states and classical random oracle}\label{sec:new-P}

In this section, we present new schemes for the task of quantum position verification. Taking cue from known protocols discussed in Section~\ref{sec:QPV}, we would like to have a protocol in which the operations to be performed by an honest prover would require practically large amount of EPR pairs to be shared between them for any be simulated by the colluding adversaries employing best known teleportation-based attacks \cite{BK11}. Here, we present 1-D quantum position verification protocols $\tn{QPV}[\tn{Ver(2),\tn{P}(\textsf{pos}),\tn{Item},\tn{Opn},\tn{Rslt}},d]$ where we make use of classical random oracle accessible to all involved parties.

\subsection{One-qubit protocol with a classical random oracle}\label{sec:P-I''}

This is similar to the protocol in Section~\ref{sec:P-I'} but with additional (classical) bits of information. This scheme is a variant of a protocol discussed in Ref.\ \cite{Unr14}. Each  party has access to a classical \textit{random} oracle,
\be\label{eqor} f : \{ 0,1 \}^{2n} \to \{ 0,1 \}~.\ee
To verify the position of $P$,  the following scheme is used.
\begin{enumerate}
	\item The verifiers agree on random $x \in \{ 0,1 \}$, and random $n$-bit strings $\bm{\theta}_0, \bm{\theta}_1 \in \{ 0,1 \}^n$, which can be also viewed as $n$-binary-digit numbers, $\bm{\theta}_0, \bm{\theta}_1 \in \{ 0,1,\dots, 2^n-1 \}$, where the $n$-bit strings represent their binary expansion. $V_0$ prepares a qubit in the state
	\be |\psi \rangle = H^{w} |x\rangle \ , \ \ w = f(\bm{\theta}_0 ,\bm{\theta}_1)\ee
	and sends it to $P$, along with $\bm{\theta}_0$. $V_1$ sends $\bm{\theta}_1$ to $P$, so they arrive at $P$ at the same time.
	\item As soon as $|\psi\rangle$ and $\bm{\theta}_0,\bm{\theta}_1$ arrive, $P$ computes (classically) $w = f(\bm{\theta}_0 ,\bm{\theta}_1)$, performs a measurement in the basis $\{ H^w |0\rangle , H^w |1\rangle \}$, and sends the outcome $x^\prime$ to both $V_0$ and $V_1$.
	\item If the verifiers receive $x^\prime$ at the time consistent with the position of $P$, \emph{and} $x^\prime=x$, then they accept; otherwise they reject.
\end{enumerate}
Even though only a single qubit is needed to run this protocol, is it not simple to break its security, even though the adversaries also have access to the classical random oracle. This is due to the fact that for the oracle to be useful, both strings $\bm{\theta}_0,\bm{\theta}_1$ are needed to be acted upon simultaneously by the oracle. To compensate, it appears that the adversaries need an exponentially large ($2^n$) entangled pairs. 

Suppose the adversaries share $2^n$ maximally entangled pairs, i.e. EPR pairs, labeled as $\bm{a}\in\{0,1,\dotsc,2^n-1\}$, each in the Bell state \eqref{eqBell}. Notice that $\bm{a}$ can be written as an $n$-digit number in binary notation ($\bm{a} \in \{ 0 , 1 \}^n$). Then they can fake $P$ following these steps:
\begin{enumerate}
	\item Upon receiving $\ket{\psi}$ and $\bm{\theta}_0$, $E_0$ teleports $\ket{\psi}$ to $E_1$ using the EPR pair labeled $\bm{a} = \bm{\theta}_0$. In doing so, $E_0$ performs a Bell measurement with outcome $k=\overline{k_0k_1}$, in binary notation. The state $E_1$ receives (instantaneously) is given by \eqref{eq:Bellmod}.
	 $E_0$ sends the result $k$ to $E_1$.
	\item At the same time, $E_1$, knowing $\bm{\theta}_1$, measures $H^{f(\bm{a}, \bm{\theta}_1)} Z H^{f(\bm{a} , \bm{\theta}_1)}$ on qubits belonging to $\bm{a}$-labeled EPR pairs. $E_1$ obtains outcomes $\bm{\lambda} = (\lambda_0,\lambda_1,\dots)$ with $\lambda_{\bm{\theta}_0} = (-1)^{x\oplus k_w}$ on the qubit that $E_0$ used to perform teleportation. $E_1$ immediately sends $(\bm{\theta}_1,\bm{\lambda})$ to $E_0$. 
	\item Upon receiving the classical information $(\bm{\theta}_1,\bm{\lambda})$ from $E_1$, $E_0$ computes $w=f(\bm{\theta}_0, \bm{\theta}_1)$, and $\lambda_{\bm{\theta}_0} (-1)^{k_w} = (-1)^{x}$ to determine $x$, and sends $x$ to $V_0$. $E_0$ knows which qubits the components of $\bm{\lambda}$ correspond to, because they are determined by $\bm{\theta}_0$. 
	\item Upon receiving the classical information $k$ and $\bm{\theta}_0$ from $E_0$, $E_1$ computes $w=f(\bm{\theta}_0 , \bm{\theta}_1)$ and multiplies $(-1)^{x\oplus k_w}(-1)^{k_w}=(-1)^x$ to determine $x$, and sends $x$ to $V_1$. Again, $E_1$ knows which qubits the components of $\bm{\lambda}$ correspond to, because they are determined by $\bm{\theta}_0$, which she just received.
\end{enumerate}

If $E_0$ and $E_1$ share $m$ EPR pairs between them, and $0\leq m\leq 2^n$, then their exists a scheme that gives lower bound on the probability with which they can succeed in cheating verifiers to be $\max\{\frac{m}{2^n},\cos^2\frac{\pi}{8}\}$. This observation follows from the discussion above and in Section~\ref{sec:P-I}.

\subsection{Two-qubit protocol with classical random oracle}\label{sec:P-II''-break}

We allow each party to have access to the classical random oracle \eqref{eqor}. The steps of this protocol are as follows:
\begin{enumerate}
	\item The verifiers agree on random $x_0, x_1, \theta \in \{ 0,1 \}$, and $n$-digit random numbers $\bm{y}_0, \bm{y}_1 \in \{ 0,1 \}^n$. $V_i$, for all $i\in\{0,1\}$, prepares a qubit in the state
	$H^\theta |x_i\rangle$,
	where $H$ is the Hadamard matrix, and sends it to $P$, along with $\bm{y}_i$ ($i\in\{0,1\}$). Here, $\tn{itm}_i=\{\bm{y}_i,H^\theta |x_i\rangle\}$ for $i\in\{0,1\}$. Both states as well as the classical information arrive at $P$ at the same time.
	\item $P$ computes (classically) $w = f(\bm{y}_0, \bm{y}_1)$, and applies $H^w$ to each of the states he received. Then he performs a Bell measurement projecting onto the state $\ket{\Psi^+}$ \eqref{eq:Bell}. If the measurement is successful, then he sends $z=1$, otherwise he sends $z=0$.
	\item The verifiers accept if the result $z$ of $P$'s measurement is consistent with the states sent by them to $P$. 
\end{enumerate}


\subsubsection*{Scheme to break the security of the protocol}\label{sec:P-varkappa-break}
A small number of pairs of entangled qubits shared by the adversaries will not break the security. In fact, it appears that the adversaries need an exponentially large number of entangled pairs. Suppose that the adversaries share $2^{n+1}+1$ maximally entangled pairs labeled as $a=0$, and $(b_0,\bm{b}_1)$, where ${b}_0 \in \{ 0,1\}$, $\bm{b}_1 \in \{ 0,1,\dotsc,2^{n}-1 \} $,  each in the Bell state \eqref{eqBell}. Then they can fake $P$ following these steps:
\begin{enumerate}
	\item Upon receiving $H^\theta \ket{x_0}$ and $\bm{y}_0$, $E_0$  teleports the state to $E_1$ using the EPR pair labeled ${a}=0$. In doing so, $E_0$ performs a Bell measurement, and $E_1$ receives the state
	\be  X^{k_0} Z^{k_1} H^\theta \ket{x_0} ~. \ee She also sends the classical information $k=\overline{k_0k_1}$, as well as $\bm{y}_0$ to $E_1$.
	\item Upon receiving $H^\theta \ket{x_1}$ and $\bm{y}_1$, $E_1$  teleports the state to $E_0$ using the EPR pair labeled $( b_0,\bm{b}_1)=(0,\bm{y}_1)$. $E_1$ receives
	\be X^{k_0'} Z^{k_1'} H^\theta \ket{x_1} ~. \ee
	She also teleports back to $E_0$ the state \eqref{eq:Bellmod2} using the EPR pair labeled $(b_0,\bm{b}_1)=(1,\bm{y}_1)$. 
	Thus, $E_0$ receives the state $X^{k_0+k_0''} Z^{k_1+k_1''} H^\theta \ket{x_0}$, which can be simplified, if $E_0$ applies $X^{k_0} Z^{k_1}$ (since $k$ is known to $E_0$) to 
	\be X^{k_0''} Z^{k_1''} H^\theta \ket{x_0}~. \ee She also sends the classical information $\bm{y}_1$ to $E_0$.
	\item $E_0$ computes $f(\bm{y}_0,\bm{b}_1)$ classically and applies $H^{f(\bm{y}_0,\bm{b}_1)}$ to each of the $(b_0,\bm{b}_1)$ channels, for $b_0=0,1$, thus effectively applying the desired $H^{f(\bm{y}_0,\bm{y}_1)}$ to the states she received from $E_1$. She then performs a Bell measurement on each of the pairs labeled $(b_0,\bm{b}_1)$, $b_0=0,1$. For each value of $\bm{b}_1$, she reports success to $E_1$, if the outcome is $\ket{\Psi^+}$ \eqref{eq:Bell}.
	\item Upon receiving $\bm{y}_0$ and the ``success" report from $E_0$, $E_1$ reports $z=0$ or $1$ to $V_1$, accordingly, knowing which pair of channels contains the teleported states. At the same time, upon receiving $\bm{y}_1$ from $E_1$, $E_0$ learns the pair of channels containing the teleported states, and reports $z=0$ or $1$, accordingly, to $V_0$. 
\end{enumerate}
The above protocol can only succeed if $k' = k'' =0$, which occurs with probability $\frac{1}{16}$. The adversaries can increase their odds at the expense of adding EPR pairs. An exponentially large number of EPR pairs are needed for 100 \% success rate \cite{BCF+14}.
Thus, the security of the protocol is breached with a number of EPR pairs shared by the adversaries that grows exponentially with the number of classical bits used in the classical oracle $f$. It is remarkable that the amount of quantum resources needed by the adversaries in this strategy grows exponentially with the length of the classical information, while the verifiers only need two independent qubits for their protocol. 

It should be pointed out that it follows from the discussion in Refs.~\cite{BFSS13,Spe16} that whenever the function $f$ (classical oracle) parametrizing the protocol can be computed by a Turing machine using logarithmic space, the adversaries can attack these protocols using EPR pairs whose number grows polynomially with the number of classical bits in $f$. Hence, it is crucial to ascertain that $f$ is a \emph{random} oracle. 

\bigskip
\section{Conclusion}
\label{sec:con}
We introduced new schemes for quantum position verification protocols by introducing a classical random oracle. We discussed the strategy for security breach by the adversaries sharing EPR pairs based on currently best known teleportation-based attacks. It is known that the entanglement distribution over long distances and the storage of entangled qubits are technologically challenging \cite{DBC99,DKD17}. The interaction between the quantum system and the environment can cause loss of information as a result of decoherence, dissipation, or decay phenomena \cite{Car09,Riv11,Wei12,DKSW18}. Quantum  memories are essential to overcome such losses caused by the environment for the preservation of the entanglement between the quantum systems for the duration longer than the decoherence period \cite{SAA+10,Riv11,DKD17}. We showed that while the verifiers need to make use of only one or two independent qubits for the verification task, the adversaries need an exponential amount of EPR pairs, depending on the number of classical bits that the verifiers make use of. In this sense, we state that the schemes we have presented are technologically feasible and practically secure.

Finally, we emphasize that our protocol is a variant of the protocol presented in Ref.\ \cite{Unr14}. The novelty of our approach is due to its practicality. It only requires a small amount of quantum resources (qubits), yet its security can only be broken by a prohibitively (at least for current technology) large amount of quantum resources. Further work is needed to provide a formal security proof. Coming up with adversarial strategies to break the protocol with as small an amount of quantum resources as possible is challenging. This makes proving security of QPV schemes, such as the one presented here, a highly non-trivial task. At the same time, it makes our results interesting to those implementing such protocols in practice.

\acknowledgments{
	We gratefully acknowledge insightful discussions with Sumeet Khatri, Bing Qi and Mark M. Wilde. We are also thankful to Fr\'ed\'eric Grosshans, Sumeet Khatri, and Florian Speelman for providing useful feedback on the manuscript. S.D.\ acknowledges support from the LSU Graduate School Economic Development Assistantship. G.S.\ acknowledges support from the U.S.\ Office of Naval Research under award number N00014-15-1-2646, the U.S.\ Army Research Office under award number W911NF-19-1-0397, and the National Science Foundation under award number OMA-1937008.
	
}

\end{document}